\documentclass[11pt]{article} 
\usepackage{hyperref} 
\pdfoutput=1 
 
\begin{document} 
 
\title{Multiple Coexisting States of a Falling Viscous Thread} 
 
\author{Mehdi Habibi, Yasser Rahmani\\
\\\vspace{6pt} Institute for Advanced Studies in Basic Sciences,\\
Zanjan, Iran\\
\vspace{12pt}
\\\vspace{6pt} Daniel Bonn\\
\\\vspace{6pt} Laboratoire de Physique Statistique,\\
Ecole Normale Sup\'erieure, Paris, France\\
\vspace{12pt}
\\\vspace{6pt} Neil M. Ribe\\
\\\vspace{6pt} Laboratoire FAST, Orsay, France
}
 

\maketitle 
 
 
\begin{abstract} 
This article describes a fluid dynamics video submission for 
the 2009 Gallery of Fluid Motion, Division of Fluid Dynamics, 
American Physical Society. 
\end{abstract} 
 
 
\section{Introduction} 
 
Two versions of the video have been submitted: a small one for use on the DFD website 
\href{http://ecommons.library.cornell.edu/bitstream/1813/14076/3/habibi_dfd09_website.mp4}{website-video} and a larger one for display at the 2009 DFD meeting
\href{http://ecommons.library.cornell.edu/bitstream/1813/14076/2/habibi_dfd09_meeting.m4v}{meeting-video}. 
 
This fluid dynamics video shows experiments in which a thin thread of 
viscous fluid (viscosity 950 cS) ejected at a rate 0.19 ml/s from a hole 2 mm in diameter falls a distance 14 cm onto a solid plate. The thread exhibits
three different states that succeed each other in time in a random manner: (1) axisymmetric stagnation flow; (2) steady coiling; and (3) rotatory folding. Transitions among these states are caused by spontaneous finite-amplitude perturbations that travel down the thread. 

\end{document}